\documentclass[aps,reprint,twocolumn,groupedaddress]{revtex4-1}
\usepackage{mathtools}
\usepackage{bm}
\usepackage{amsmath,amssymb,graphicx}
\usepackage{mathtools}
\usepackage{bm}
\usepackage{color}
\usepackage{braket}
\usepackage{textgreek}
\usepackage{graphicx}
\usepackage[percent]{overpic}
\usepackage{xcolor}
\usepackage{upgreek}

\begin{document}

\title{Control of spontaneous emission dynamics in microcavities with chiral exceptional surfaces}

\author{ Q. Zhong,$^{1}$ A. Hashemi,$^{1}$ {\c S}. K. {\"O}zdemir,$^{2,3}$ and R. El-Ganainy$^{1,4,}$ }

\email[Corresponding author: ]{ganainy@mtu.edu}

\affiliation{$^1$Department of Physics, Michigan Technological University, Houghton, Michigan 49931, USA}

\affiliation{$^2$Department of Engineering Science and Mechanics, The Pennsylvania State University, University Park, Pennsylvania 16802, USA}

\affiliation{$^3$Materials Research Institute, The Pennsylvania State University, University Park, Pennsylvania 16802, USA}

\affiliation{$^4$Henes Center for Quantum Phenomena, Michigan Technological University, Houghton, Michigan 49931, USA}

\begin{abstract}
We investigate spontaneous emission from a quantum emitter located  within the mode volume of a microring resonator  that features   chiral  exceptional points. We show that this configuration offers enough degrees of freedom to exhibit a full control to either enhance or suppress the emission process. Particularly, we demonstrate that the Purcell factor can be enhanced by a factor of two beyond its value in an identical microring operating at a diabolic point. Our conclusions, which are derived using a non-Hermitian Hamiltonian formalism, are confirmed by employing full-wave simulations of realistic photonic structures and materials. Our results offer a straightforward route to improve the performance of single photon sources using current photonics technology without the need for building optical resonators with ultrahigh quality factors or nanoscale volumes.


\end{abstract}


\maketitle

\section{Introduction}
 Quantum engineering seeks to utilize quantum mechanics to build a new generation of computing machines, encryption schemes, and sensing devices with unprecedented performance in terms of computational power, security strength, and sensitivity, among other applications. In the pursuit to achieve these goals, several material platforms provide complementary solutions to overcome various practical hurdles. These include trapped atoms \cite{Gross995}, superconducting circuits \cite{Blais2020}, and photonics \cite{Brien1567}. The latter is particularly interesting due to its mature technology and natural interface with current optical communication systems. At the heart of modern quantum optics technology is the ability to control light-matter interaction at the quantum level for various applications such as building nonclassical light sources \cite{Zeilinger_2017}, optical transistors \cite{Chang2007}, and quantum memory \cite{Lvovsky2009}. In this regard, efforts have been recently dedicated to building efficient single photon sources that can produce individual photons on demand at high repetition rates \cite{Vucovic2012}. This progress was enabled by engineering various optical resonator geometries that can support small modal volumes and large quality factors to tailor the photonic local density of states (PLDOS) surrounding quantum emitters (QE), hence controlling their spontaneous emission (SE) rates as quantified by the Purcell factor (PF) \cite{Purcell1946} (see Ref.\:\cite{Hiroyuki2019} for detailed discussions). Examples include planar photonic crystals, vertical Bragg reflectors,  microdisks and plasmonic structures. For comprehensive reviews and performance comparison, see Ref.\:\cite{Vahala2003}. Despite these promising results, the aforementioned arrangements are not easy to mass produce or integrate with other photonics components. An attractive alternative in terms of mass fabrication and large-scale integration is microring resonators \cite{vahala2004}. On the downside, however, microring resonators suffer from relatively large mode volumes and limited quality factors \cite{Vahala2003}. It will be thus of interest to devise new routes for controlling SE in microring resonators and possibly enhance their PF beyond their current performance.

Motivated by this goal, here, we study the interaction between light and a QE in a family of microring resonators whose design is tailored to operate in the vicinity of or at a special type of non-Hermitian singularities known as chiral exceptional points (EPs). At an EP, two or more eigenmodes of a system fuse together, leading to a reduction of the dimensionality associated with the eigenspace of the system (for recent reviews on the physics of non-Hermitian optical systems and EPs see Refs.\:\cite{ElGanainy2018PRL, Feng2017JMP, Ozdemir2019JPL, Miri2019JPA}). In contrast to recent investigations of SE near isolated EPs \cite{PhysRevLett.117.107402, Pick:17, PhysRevB.96.224303,Hadad5576, Wiersig2020PRR}, here instead, we consider a system that features an exceptional surface \cite{Zhong2019SES,Zhong:19,ElGanainy2020PRApp} which offers robust operation and tunability. As we will see, this geometry offers several advantages in terms of controlling PF [by suppressing it completely or enhancing its value by a factor of two compared with microring resonators operating at diabolic points (DPs)] and integration with a waveguide to collect the emitted photons from a predetermined port.\\

\section{Formalism}
\begin{figure}
\begin{center}
\includegraphics[width=3in]{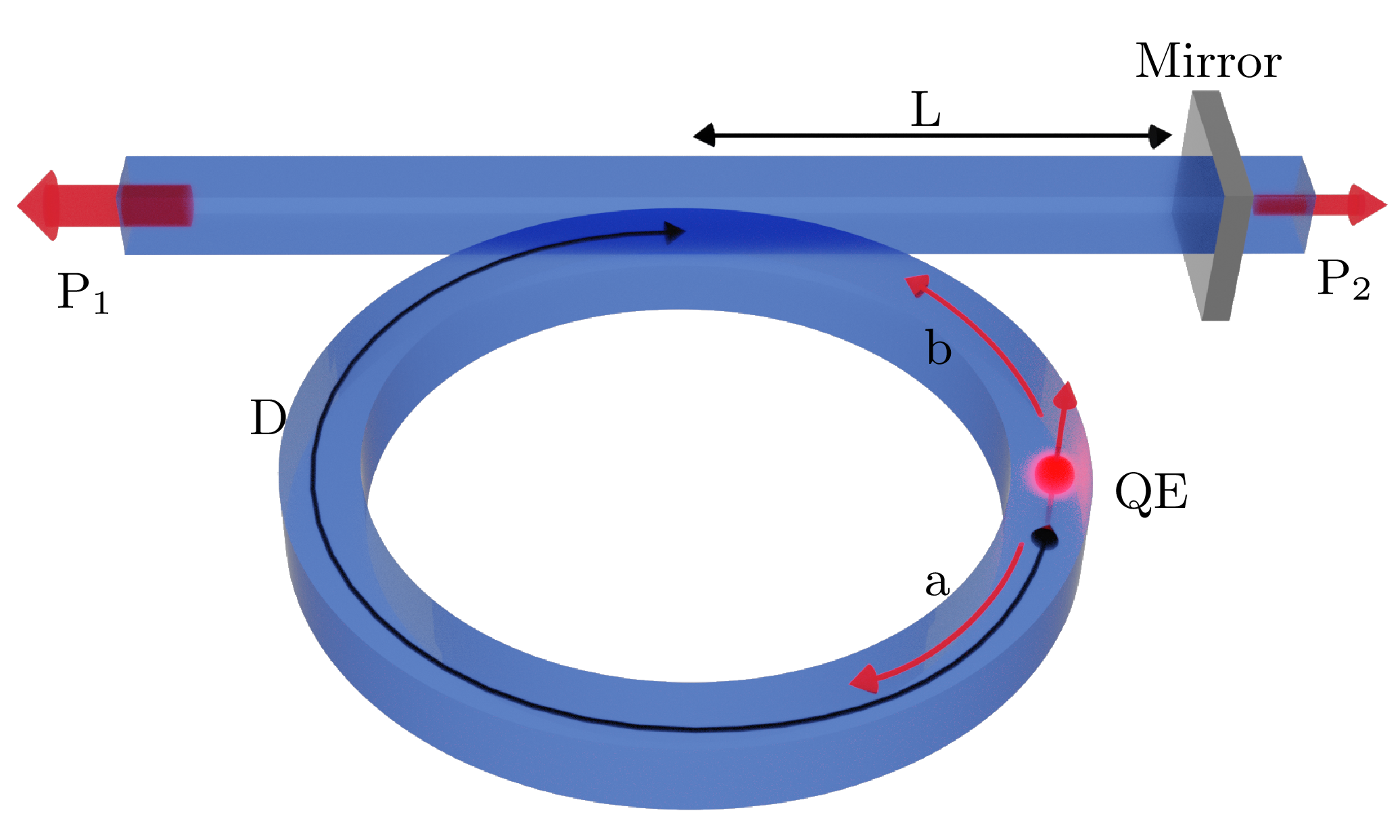}
\caption{A schematic of the proposed geometry. It consists of a microring resonator evanescently coupled to a waveguide with an end mirror as proposed in Ref.\:\cite{Zhong2019SES} for implementing exceptional surfaces. Here, we assume that a quantum emitter (QE) is located inside the microring resonator. Under resonant conditions (transition frequency between the QE energy levels matches the eigenfrequencies of the optical modes of the ring resonator), the QE couples to the clockwise (CW) and counterclockwise (CCW) modes of the ring. Photons emitted from the QE take different paths and self-interfere before they arrive at the exit port (P$_1$ in the figure). This can lead to the enhancement or suppression of Purcell factor (PF).}
\label{Fig:Schematic}
\end{center}
\end{figure}

 The optical platform we consider in this study is shown in Fig.\:\ref{Fig:Schematic}. It consists of a microring resonator coupled to a waveguide terminated by a mirror at one of its ports. In the absence of the mirror, the optical resonator has two degenerate modes, i.e., it operates at a DP. On the other hand, the mirror provides unidirectional coupling between the clockwise (CW) and counterclockwise (CCW) modes and, hence, induces a chiral EP (in fact an exceptional surface). This structure was recently proposed as a robust implementation for EP-based optical sensors \cite{Zhong2019SES}, optical amplifiers \cite{ElGanainy2020PRApp}, and directional absorbers \cite{Zhong:19}. In these previous papers, the system was externally excited through the waveguide, and the response was studied by monitoring the transmission or reflection spectra under different conditions. Here, we study the emission properties of a QE located within the mode volume of the resonator (Fig.\:\ref{Fig:Schematic}) by monitoring the emitted optical power collected by the waveguide ports. We are particularly interested in evaluating how the presence of this chiral EP modifies the PF of the QE, i.e., the decay rate from the excited state to the ground state in the absence of any external active perturbation (no applied optical or electrical signal).

To simplify the analysis and gain insight into the problem, we will employ a technique based on a non-Hermitian Hamiltonian $\hat{H}=\hat{H}_R+\hat{H}_E+\hat{H}_I$ (with the sub-indices $R, E$, and $I$ referring to radiation, emitter and interaction, respectively) and neglect the effect of quantum jumps. The system shown in Fig.\:\ref{Fig:Schematic} can be represented by the Hamiltonian
\begin{equation} \label{Detailed_H}
\begin{aligned}
\hat{H}_R &=\hbar (\omega_o -i\gamma_W) (\hat{a}^\dagger \hat{a} + \hat{b}^\dagger \hat{b})+\hbar \kappa \hat{a} \hat{b}^\dagger, \\
\hat{H}_E &=\hbar (\omega_e-i\gamma_e) \ket{e}\bra{e}, \\
\hat{H}_I &=\hbar (J_\text{CW }\hat{a}^\dagger + J_\text{CCW} \hat{b}^\dagger) \ket{g}\bra{e}+\text{H.c.}
\end{aligned}
\end{equation}
In the radiation Hamiltonian $\hat{H}_R$, $\hat{a}^\dagger$, $\hat{a}$,  and $\hat{b}^\dagger$, $\hat{b}$ are the creation and annihilation operators of the CW and CCW optical modes, which have the same resonant frequency $\omega_o$ and loss rate $\gamma_W$ (coupling loss). The unidirectional coupling from the CW to the CCW mode is   denoted by $\kappa =-2i\gamma_W |r|e^{i\phi_W}$ \cite{Zhong2019SES, vahala2004}.  Here, $|r|$ is the field reflection amplitude from the mirror, and $\phi_W=2 \beta L+\phi_r$, where $\beta$ is the propagation constant in the waveguide, $L$ is the distance between the waveguide-ring junction and the mirror, and $\phi_r$ is the field reflection phase from the mirror (i.e., $r=|r|e^{i\phi_r}$). Note that $\hat{H}_R$ (which can be inferred from Eq.\:(1) in \cite{Zhong2019SES} by elevating the classical field variables into operators) is not Hermitian. In general, the Hamiltonian $\hat{H}_R$ will exhibit an EP of order two in the single-photon  subspace. In the emitter Hamiltonian $\hat{H}_E$, $\omega_e$ denotes the transition frequency between the ground state $\ket{g}$ and excited state $\ket{e}$, and $\gamma_e$ quantifies the excited state lifetime due to coupling only to free space continuum modes. We   assume that the emitter does not radiate efficiently into free space directly, which we will justify later when discussing implementations. In the interaction Hamiltonian $\hat{H}_I$,  $J_{\text{CW,CCW}}=\vec{\mu}\cdot \vec{E}_{\text{CW,CCW}}(\vec{r})$ are the coupling constants between the CW and CCW optical modes and an emitter having an electric dipole moment $\vec{\mu}$ and located at $\vec{r}$, and $\vec{E}_{\text{CW,CCW}}(\vec{r})$ are the normalized electric fields of CW and CCW modes of the bare microring resonator (i.e., without mirror or QE), which we use as bases for expanding the optical fields at the location of the emitter $\vec{r}$ \cite{Charles2010}.  For these CW/CCW modes, the magnitude of the electric field at a particular transverse position does not vary along the angular direction. Instead, the field just acquires a phase. If we chose the ring-waveguide junction as a reference point, we can then write $J_{\text{CW}}=Je^{-i\phi_E}$, and $J_{\text{CCW}}=Je^{i\phi_E}$, where $\phi_E=\beta D$ (see Fig.\:\ref{Fig:Schematic}), and H.c. stands for Hermitian conjugate.

Within the single excitation subspace, which is relevant to the SE process, the general wavefunction can be written as $\ket{\psi(t)}=a(t) \ket{1,0,g}+b(t) \ket{0,1,g}+c(t) \ket{0,0,e}$, where the coefficients $a(t)$, $b(t)$, and $c(t)$ are the probability amplitudes of finding the excitation either in the CW mode, CCW mode, or in the QE. Importantly, it is straightforward to show that $[\hat{N},\hat{H}]=0$, where $\hat{N}=\hat{a}^\dagger \hat{a} + \hat{b}^\dagger \hat{b} + \ket{e}\bra{e}$ is a generalized number operator that accounts for the total excitations in the bosonic modes and the QE. In other words, the Hamiltonian $\hat{H}$  conserves the number of excitations. However, as a result of its non-Hermitian character, it does not conserve the probability of finding the excitation trapped inside the system, i.e. $p(t)=|a(t)|^2+|b(t)|^2+|c(t)|^2 \leq 1$. By substituting $\psi(t)$ in the Schr\"odinger's equations $i\hbar d\ket{\psi(t)}/dt=\hat{H}\ket{\psi(t)}$ and projecting on the states $\ket{1,0,g}$, $\ket{0,1,g}$, and $\ket{0,0,e}$, we obtain
\begin{equation} \label{Eq:CMT}
i\frac{d\vec{v}}{dt}=H \vec{v}, 
H=
\begin{pmatrix}
\omega_o-i\gamma_W & 0 & Je^{-i\phi_E} \\
\kappa & \omega_o-i\gamma_W & Je^{i\phi_E} \\
Je^{i\phi_E} & Je^{-i\phi_E} & \omega_e-i\gamma_e
\end{pmatrix},
\end{equation}
where $\vec{v}=\left(a(t), b(t), c(t)\right)^T$, with the superscript $T$ indicating matrix transpose. Note that, consistent with the expression for $\hat{H}$, the effective discrete Hamiltonian $H$ is non-Hermitian and exhibits an exceptional surface when $J=0$. To study SE using Eq.\:(\ref{Eq:CMT}), we consider the initial condition $a(0)=b(0)=0$ and $c(0)=1$, i.e., the emitter is initially in the excited state, and the optical modes are in the vacuum state.

\section{Results}
 To ensure that the system operates in the vicinity of the EP, we consider the weak coupling regime $J\ll \gamma_W$ with negligible Markovian effects. In other words, a photon emitted from the QE into the photonic mode will escape quickly to the waveguide environment before it is able to couple back to the QE. In microring resonators with large mode volume, this assumption is valid. In addition, for resonators having quality factors in the order of $Q \sim 5000$, the photon lifetime is $2Q/\omega_o \sim 10^{-1} \: \text{ns}$.  Meanwhile, the typical lifetime of the excited state associated with a quantum dot in free space is $\sim$ 1--10 ns. In this regime of operation, one can approximate the decay into the two channels as independent processes. This is mainly justified by the absence of interference between these two processes. In Appendix A,  we discuss the validity of this approximation. 

Under these conditions, we can integrate the last equation in Eq.\:(\ref{Eq:CMT}) and obtain $c(t) \sim e^{-i \omega_e t}$. By substituting back in the first two equations (which is equivalent to performing adiabatic elimination), we obtain
\begin{equation} \label{Eq:CMT_weak}
i\frac{d}{dt}
\begin{pmatrix}
a  \\
b
\end{pmatrix}
=
\begin{pmatrix}
\omega_o-i\gamma_W & 0 \\
\kappa & \omega_o-i\gamma_W\\
\end{pmatrix}
\begin{pmatrix}
a  \\
b
\end{pmatrix}
+ J
\begin{pmatrix}
e^{-i (\omega_e t+\phi_E)} \\
e^{-i (\omega_e t-\phi_E)}
\end{pmatrix}.
\end{equation}
By seeking a solution of the form $\left(a(t), b(t)\right)^T= \left(A,B\right)^T e^{-i \omega_e t}$, we find the steady state solution as
\begin{equation} \label{Eq:A_B}
\begin{aligned}
A=&\frac{Je^{-i\phi_E}}{\Delta+i\gamma_W}, \\
B=&\frac{Je^{i\phi_E}}{\Delta+i\gamma_W}+\frac{J \kappa e^{-i\phi_E}}{\left(\Delta+i\gamma_W \right)^2},
\end{aligned}
\end{equation}
where $\Delta=\omega_e-\omega_o$. As expected, the $\kappa$ term, which is responsible for the existence of the exceptional surface when $J=0$, introduces a second-order pole in the expression for $B$. From the above expressions, we observe that, for $J\ll \gamma_W$, $|A| \& |B|\ll 1$ (even when $\Delta=0$), which justifies the weak coupling approximation used to arrive at these results. From a classical perspective, one can think of $A$ and $B$ as the steady state field amplitudes of the CW and CCW modes under excitation by a driven classical dipole antenna. The output signal from ports 1 and 2 will be thus $s_{\text{EP}}^{(1)}=-\sqrt{2\gamma_W} \left( A|r|e^{i\phi_W}+B \right)$ and $s_{\text{EP}}^{(2)}=-\sqrt{2\gamma_W} \left[|t| e^{i (\beta L+\phi_t)}  A \right]$, where $\phi_t$ is the phase associated with the field transmission coefficient $t$ from the mirror, i.e, $t=|t|e^{i\phi_t}$. The total normalized power emitted from both ports $P_{\text{EP}}\equiv |s_{\text{EP}}^{(1)}|^2+|s_{\text{EP}}^{(2)}|^2$ is given by
\begin{equation} \label{Eq:Power}
{P_{\text{EP}}}= P_{\text{DP}}\frac{|\chi+|r|+e^{i \Delta \phi}|^2+|t|^2}{2},
\end{equation}
where the $P_{\text{DP}}\equiv P_{\text{EP}}|_{r=0}=\frac{4\gamma_W J^2}{\Delta^2+\gamma_W^2}$ is the power emitted in the DP case. In Eq.\:(\ref{Eq:Power}), $\Delta \phi \equiv 2\phi_E-\phi_W$ and $\chi \equiv \frac{\kappa e^{-i \phi_W}}{\Delta+i\gamma_W}  =\frac{-2i\gamma_W |r|}{\Delta+i\gamma_W} $. Note that $P_{\text{EP,DP}}$ are the power emitted only into the waveguide and do not account for coupling to free space. By taking this latter effect into consideration as an additive process and denoting the power decay PF enhancement as $\eta$, we finally arrive at \cite{Vuckovic1999}
\begin{equation} \label{Eq:eta}
\eta \equiv\frac{PF_{\text{EP}}}{PF_{\text{DP}}} = \frac{2\gamma_e+P_{\text{EP}}}{2\gamma_e+P_{\text{DP}}}=\frac{\mathcal{F}+\mathcal{R}}{\mathcal{F}+1},
\end{equation}
where $PF_{\text{EP,DP}}$ are the Purcell enhancement factors  in the EP and DP cases with respect to emission only into free space, i.e., $PF_{\text{EP,DP}}=\frac{2\gamma_e+P_\text{EP,DP}}{2\gamma_e}$. The variables $\mathcal{R}$ and $\mathcal{F}$ in Eq.\:(\ref{Eq:eta}) are defined as $\mathcal{R}\equiv \frac{P_{\text{EP}}}{P_{\text{DP}}}$ and $\mathcal{F}\equiv \frac{2\gamma_e}{P_{\text{DP}}}=\frac{1}{PF_{\text{DP}}-1}$. When $r=0$ (and $t=1$), the system operates at a DP, and hence, we have $\eta=1$, as expected. In the limit of $r=1$ (i.e., perfect mirror), the system is at the EP, and for $\Delta=0$, we have resonant emission with $\chi=-2$. Under these conditions, we have $\mathcal{R}=1-\cos\Delta \phi$, which attains its maximum value $\mathcal{R}_\text{max}=2$ at $\Delta \phi=(2m+1)\pi$ for integer $m$. For $\gamma_e\ll P_{\text{DP}}$ or equivalently $PF_{\text{DP}}\gg1$ (a valid assumption for typical Purcell enhancement settings), this translates into $\eta=2$. On the other hand, for $\Delta \phi=(2m \pm 1/2)\pi$, we obtain $\eta=1$, i.e., equivalent to the DP case at the same resonant frequency condition. Interestingly, for $\Delta \phi=2m\pi$, we find  that $\mathcal{R}=0$, and $\eta$ attains its minimum value of $\eta_{\rm min}=\frac{2\gamma_e}{2\gamma_e+P_{\text{DP}}}$, i.e., the QE radiates with a rate $2\gamma_e$ as if it was located in free space.

\begin{figure}[t]
\begin{center}
\includegraphics[width=3.4in]{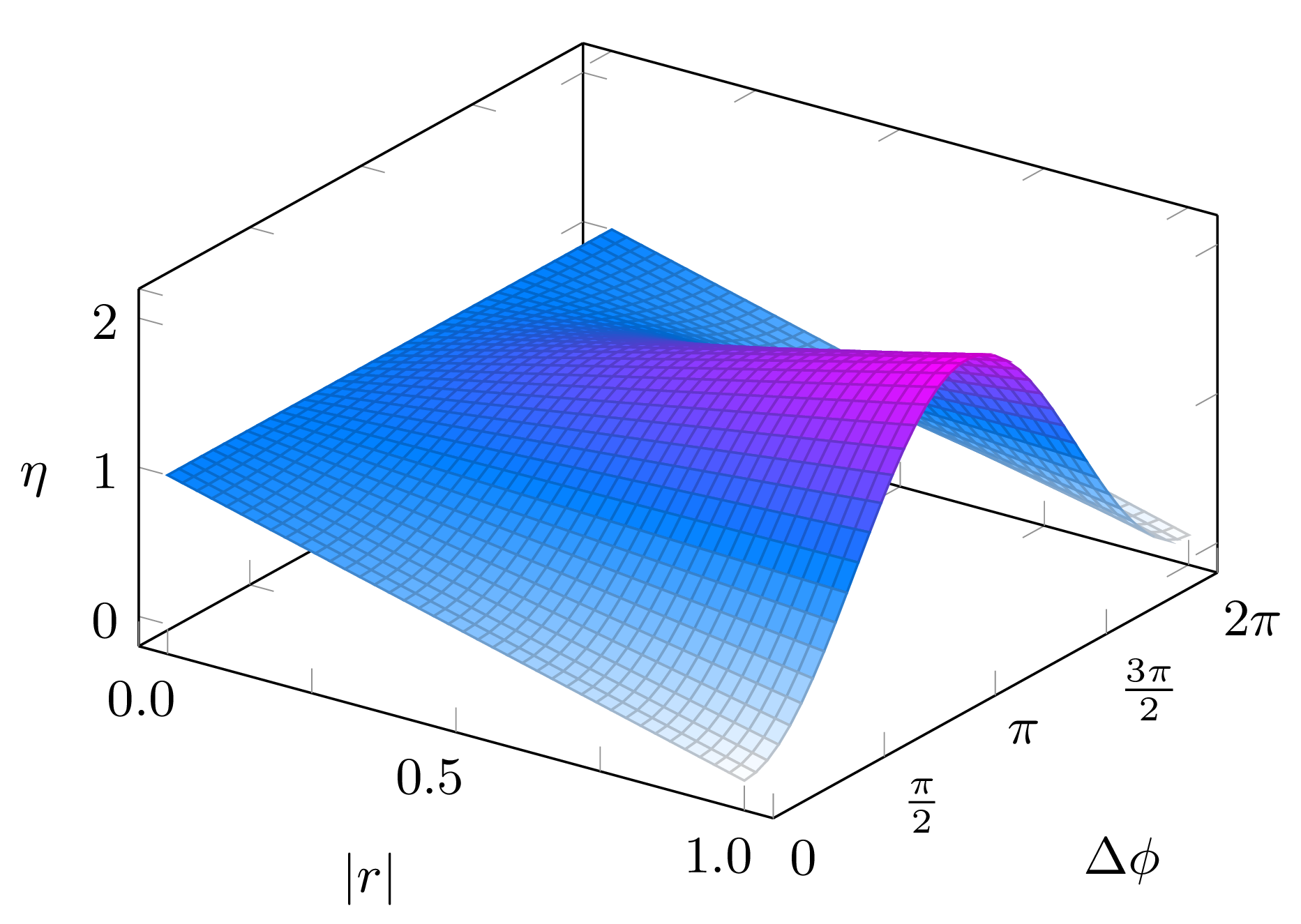}
\caption{Purcell factor enhancement $\eta$ as a function of the mirror field reflectivity amplitude $|r|$ and $\Delta \phi$ (characterizing the change in the position of the mirror or the QE) under the resonant condition $\Delta=0$. The absorption of the mirror in the above plot is assumed to be zero. The maximum enhancement  $\eta=2$  occurs at $\Delta \phi=\pi$. For more realistic mirrors with a finite absorption coefficient, the peak enhancement will be less than its maximum possible value  attainable  in this ideal case.}
\label{Fig:2Dplot}
\end{center}
\end{figure}

These results show that the platform shown in Fig.\:\ref{Fig:Schematic} provides enough degrees of freedom to tune the SE rate from that of free space emission to an enhanced SE regime with an enhancement twice that of the DP case. This is also evidenced in Fig.\:\ref{Fig:2Dplot}, which depicts $\eta$  as a function of both $|r|$ and $\Delta \phi$  under the condition $PF_{\text{DP}}\gg1$. An interesting feature of Eq.\:(\ref{Eq:CMT_weak}) can be observed by writing $B=\frac{Je^{i\phi_E}} {\left(\Delta+i\gamma_W \right)^2} \times \left(\Delta+i\gamma_W -2i\gamma_W |r| e^{-i\Delta\phi} \right)$. When $\cos\Delta\phi=1/ (2|r|)$ and $\Delta= \pm\gamma_W \sqrt{4|r|^2-1}$ (the sign depends on the value of $\Delta\phi$) with $|r|\geq 0.5$, we have $B=0$, and $\eta=0.5$. In this case, the SE rate is suppressed compared with that associated with the DP, but the emission becomes chiral with the photon emitted only in the CW mode. This feature, which arises due to destructive interference between the CCW mode and the back-reflected wave in the waveguide inside the ring, has been recently observed experimentally using PT-symmetric microwave and acoustic setups \cite{Chen2020NPhy}.

Before we proceed, it is worth mentioning that the above results can be obtained by using the notion of PLDOS. In this case, the EP associated with the photonic Hamiltonian gives rise to a double pole in the spectrum of the Green's operator. Consequently, the nonvanishing coupling between the QE and the Jordan vector yields the square Lorentzian term of Eq.\:(\ref{Eq:A_B}). Indeed, a straightforward calculation of the Green's operator associated with the Hamiltonian matrix $H$ (after eliminating the third row and column that account for the QE) yields expressions equivalent to Eq.\:(\ref{Eq:A_B}). For a more detailed mathematical discussion on how the Jordan vector modify the PLDOS and SE, see Refs.\:\cite{Pick:17,Chen2020NPhy, Wiersig2020PRR}.

\begin{figure}
\begin{center}
\includegraphics[width=3in]{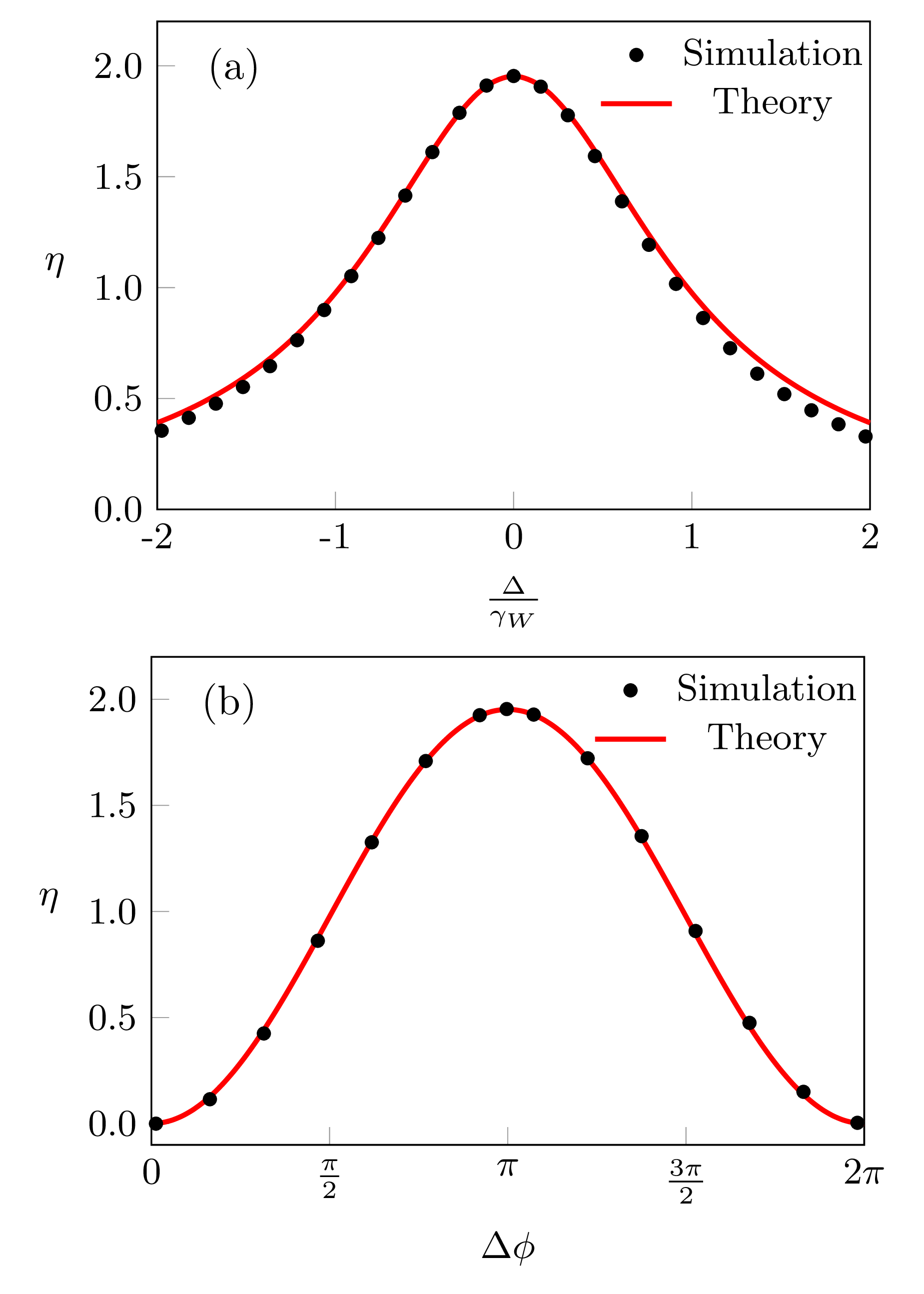}
\caption{Plots of (a) the value of Purcell factor (PF) enhancement $\eta$ as a function of the normalized frequency detuning $\frac{\Delta}{\gamma_W}$ for $\Delta \phi=\pi$, and $|r|=0.976$; and (b)  $\eta$ as a function of the mirror (or QE) position as parametrized by $\Delta\phi$, under resonant condition $\Delta=0$. In producing this plot, we used $|r|=0.976$, as obtained for our realistic implementation of the mirror. Red solid line represents the results obtained by using Eq.\:(\ref{Eq:eta}) together with the optical parameters of the device, obtained as outlined in Appendix \ref{Appendix: numerical evaluation}. Black dots represent the results obtained by full-wave simulations.}
		\label{Fig:1Dplot}
\end{center}
\end{figure}

Next, we confirm the above predictions by performing two-dimensional full-wave finite-difference time-domain (FDTD) simulations using realistic structure dimensions and material systems. In our simulations, the microring resonator has  a  refractive index of $n_r=3.47$ embedded in a background with $n_b=1.44$. The outer radius of the microring is taken to be $R=5 \: \mu\text{m}$, and its width is $w=0.25\: \mu\text{m}$. The edge-to-edge separation between the ring and the waveguide is $d=0.2\: \mu\text{m}$, and its width is identical to that of the ring waveguide. The mirror at end of the waveguide is made of a 100-nm-thick silver layer. These design parameters lead to the following optical properties  for the TE optical modes: effective refractive index $n_\text{eff}=2.93$, $\omega_o=1216 \: \text{THz}$ or equivalently $\lambda_o=1549 \: \text{nm}$, $\gamma_W=124 \: \text{GHz}$, corresponding to $Q=4900$,  and $|r|=0.976$.  Due to the optical absorption of silver, we find $|t|=0.008$. The numerical evaluation of the above parameters is presented in detail in Appendix \ref{Appendix: numerical evaluation}.
The QE in our simulations is located as shown in Fig.\:\ref{Fig:Schematic}, i.e., at one-quarter of the perimeter as measured from the ring-waveguide junction in the clockwise direction. Its dipole moment tensor is assumed to have a component perpendicular to the ring plane and, thus, couples only to the TE optical mode. In the FDTD simulations, the QE is modeled by using an oscillating classical electric dipole. Note that these calculations can provide information about the PF \cite{El_Ganainy_2013,Vuckovic1999} but not about the temporal `wavefunction' of the photon. 

Figure \ref{Fig:1Dplot}(a) plots $\eta$ as a function   of $\Delta/\gamma_W$ for the optimal mirror position that maximizes the PF for the device described above. The theory plot (red solid line) presents the results as obtained from Eq.\:(\ref{Eq:eta}) and the optical parameters of the device which are extracted from FDTD simulations  (Appendix \ref{Appendix: numerical evaluation} for details). On the other hand, the curve denoted as simulation (black dots) presents the same information as obtained directly by calculating the optical power at the output ports of the waveguide from the FDTD simulations (with and without the mirror). In both cases, we applied the definition of $\eta$ under the condition $PF_{\text{DP}}\gg1$. In our system, this approximation can be justified by noting that, in our FDTD simulations, the power emitted outside the microring resonator to free space is negligible compared with that emitted to the waveguide ports. Evidently good agreement is observed especially at the resonant frequency. Due to the finite reflectivity of the mirror, the maximum enhancement here is $\eta=1.95$. Next, we plot the values of $\eta$ as a function of $\Delta \phi$. In our simulations, we change $\Delta \phi$ by scanning the mirror position around its optimal value. Particularly, we varied the mirror position up to 440 nm with 20 nm increments. Again, we observe a good agreement between theory and numerical simulations.

\begin{figure}
\begin{center}
\includegraphics[width=3.4in]{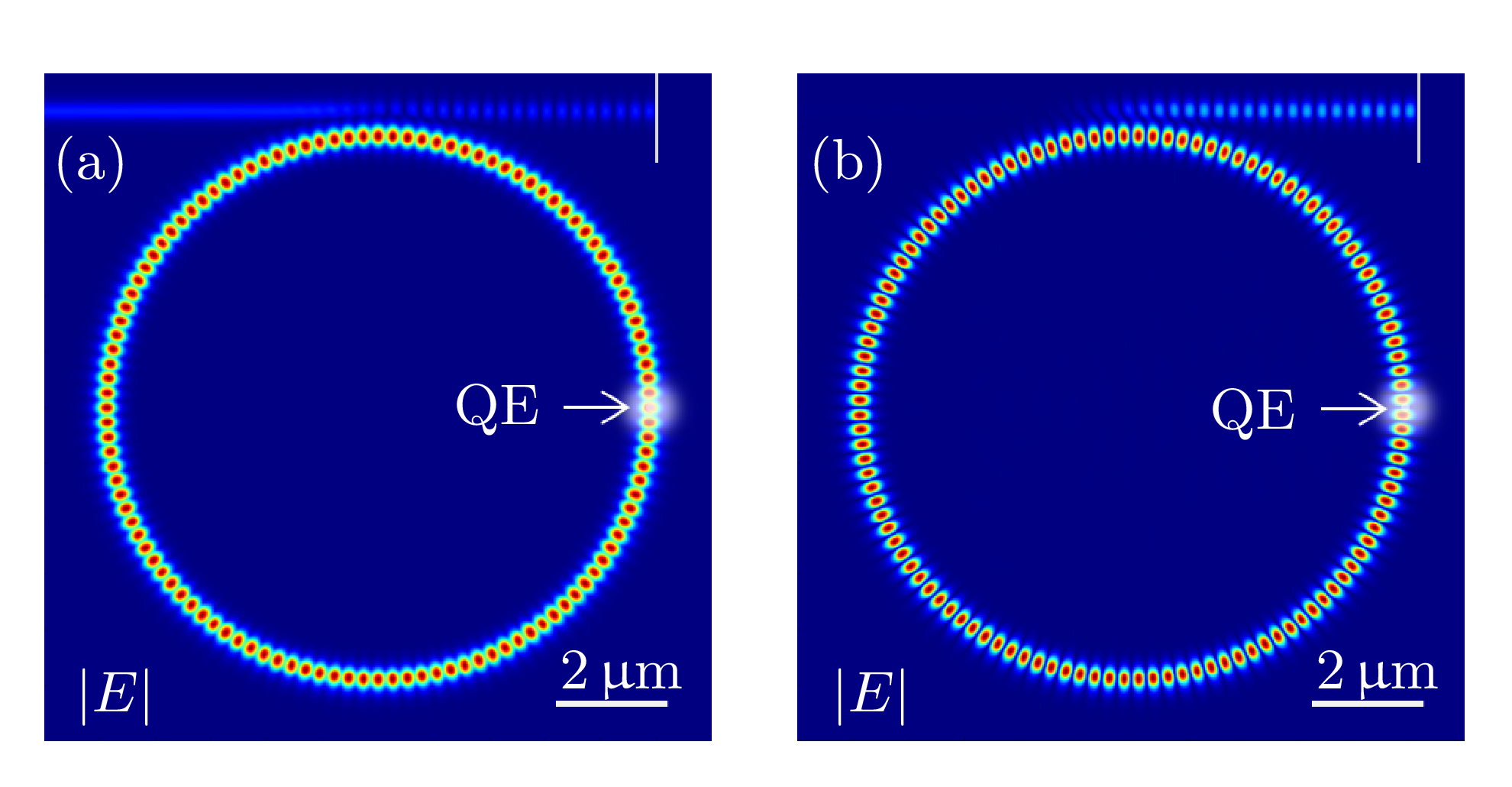}
\caption{Distribution of the electric field amplitudes in the structure under semi steady state conditions for the two different extreme scenarios: (a) $\Delta\phi=\pi$ ($L=5050$ nm) and (b) $\Delta\phi=0$ ($L=5180$ nm). In the first case, the amplitudes of the clockwise (CW) and counterclockwise (CCW) waves are different by a factor of three, larger [than the diabolic point (DP) case] power output from $\text{P}_1$, which leads to maximum Purcell factor (PF) enhancement. In the second case, the CW and CCW waves have identical amplitudes, forming a perfect standing wave pattern inside the ring, with the null located at the position of the quantum emitter (QE). This leads to a decoupling between the QE and the optical modes. As a result, the PF is suppressed ($\eta=0$) as evidenced by the absence of the optical power in the output port $\text{P}_1$.}
\label{Fig:Fields}
\end{center}
\end{figure}

To gain insight into these results, we investigate the electric field distribution associated with various mirror positions as shown in Fig.\:\ref{Fig:Fields}, focusing on the cases  when $\Delta\phi=\pi$ (optimal mirror position) and $\Delta\phi=0$ (trapping condition). In our simulations, these values correspond to $L=5050$ and $L=5180\: \text{nm}$, respectively. From Fig.\:\ref{Fig:Fields}(a), we observe that the field in the left port of the waveguide is uniform, indicating an escaping traveling wave as expected. The  field inside the ring forms an imperfect standing wave pattern (the amplitude of the CCW component three times larger than that of the CW component) having a peak at the QE location, which explains the enhancement of the PF.  On the other hand, the field distribution in Fig.\:\ref{Fig:Fields}(b) demonstrates a perfect standing wave inside the ring with a node located at the QE position. As a result, the photonic mode effectively decouples from the QE, leading to near zero PF. In this latter scenario, the microring and the right section of the waveguide with its end mirror form a resonator that traps the excitation, with the lifetime determined by the mirror reflectivity and the radiation rate from the ring to free space. \\

\section{Conclusion}
 In summary, we have investigated the interaction between light and a QE in a microring optical resonator exhibiting a chiral EP. Our analysis based on the non-Hermitian Hamiltonian approach shows that, in the weak coupling regime, the presence of an EP can enhance the  PF  by a factor of two. Furthermore, implementing the chiral EP by side-coupling the resonator to a waveguide terminated by a mirror can offer enough degrees of freedom to even significantly suppress the SE process. These conclusions are confirmed by using full-wave FDTD simulations. Our results open the door for designing more efficient single photon sources. In future papers, we intend to explore this possibility by studying the case of continuously driven QE near an EP.

\appendix
\section{Evaluation of Purcell effect via eigenvalue analysis}\label{Appendix:eigenvalue analysis}
In the main text, we used the method presented in Ref.\:\cite{Vuckovic1999} to derive the $\eta$ factor. This was done to make a connection with our computational study. In this section, we derive the same results by directly solving Eq.\:(\ref{Eq:CMT}). The general solution can be expressed as a sum of eigenvectors, each scaled by an exponential time evolution factor given by the corresponding eigenvalue of the matrix $H$. We note, however, that the Purcell regime lies in the deep weak coupling domain where the decay rate of the QE is only modified without significantly impacting the eigenvectors. Thus, one can simplify the problem by focusing only on the correction introduced to the free space decay rate $\gamma_e$ by the photonic environment. There are several equivalent techniques to do so. For instance, one can solve the linear equations using Laplace transformation and employing the so-called dominant pole approximation \cite{El_Ganainy_2013,Teimourpour_2015} or alternatively by using a perturbative expansion. Here, we use this latter technique. Particularly, by seeking a series solution for the eigenvalues of $H$ as a function of the perturbation parameter $\epsilon=J/\alpha$ with $\alpha=\gamma_W-\gamma_e$, we find that the new value for the otherwise bare (i.e., obtained with $J=0$ ) eigenvalue $\omega_e-i\gamma_e$ is now given by
\begin{equation}
\lambda_e=\omega_e-i\gamma_e-\delta+\mathcal{O}\{\epsilon^4\},
\end{equation}
where
\begin{eqnarray}
\delta =-\frac{J^2 \left(2 \Delta +\kappa  e^{-2 i \phi_E }+2 i \alpha\right)}{\left(\Delta +i \alpha\right){}^2}.
\end{eqnarray}
Note that $\delta$ is in general a complex number. The real part of $\delta$ corresponds to the Lamb shift and is negligible in the Purcell regime. The imaginary part, on the other hand, corresponds to the modification of the decay rate of the excited state amplitude. Thus, the total amplitude SE rate is now given by
\begin{eqnarray} \label{Eq:Decay_Appendix}
\gamma_t=\gamma_e+\text{Im}\{\delta\},
\end{eqnarray}
where
	\begin{equation}\label{Eq:imaginary_delta}
	\begin{split}
	\text{Im}\{\delta\}=\frac{2 J^2 }{\left(\Delta ^2+\alpha^2\right){}^2} \left[|r| \gamma_W\cos (\Delta \phi )(\Delta ^2-\alpha^2) \right. \\
	 \left. -2 \Delta  |r|\gamma_W \alpha\sin (\Delta \phi )+\alpha(\Delta ^2+\alpha^2)\right].
	\end{split}
	\end{equation}
Equation (\ref{Eq:Decay_Appendix}) indeed confirms the validity of our approximation of treating the two decay processes (into free space and the waveguide) as two additive independent effects. Moreover, one can quickly confirm the validity of the expression in Eq.\:(\ref{Eq:imaginary_delta}) in some simple scenarios. For instance, for the resonant DP case ($r=0$ and $\Delta=0$), Im\{$\delta$\}$=2J^2/\alpha$, which corresponds to the expression in the main text when $\gamma_e \ll\gamma_W$. On the other hand, for the EP case with $|r|=1$, $\Delta=0$, and $\Delta\phi=\pi$, we find Im\{$\delta$\}$=2J^2(\alpha+\gamma_W)/\alpha^2$ also consistent with our results in the main text for $\gamma_e \ll\gamma_W$. Finally, for $|r|=1$, $\Delta=0$,  and $\Delta\phi=0$, we obtain Im\{$\delta$\}$=2J^2(\alpha-\gamma_W)/\alpha^2 $ $\rightarrow$ 0 when $\gamma_e \ll \gamma_W$, as expected.

\begin{figure}[t]
	\begin{center}
		\includegraphics[scale=1]{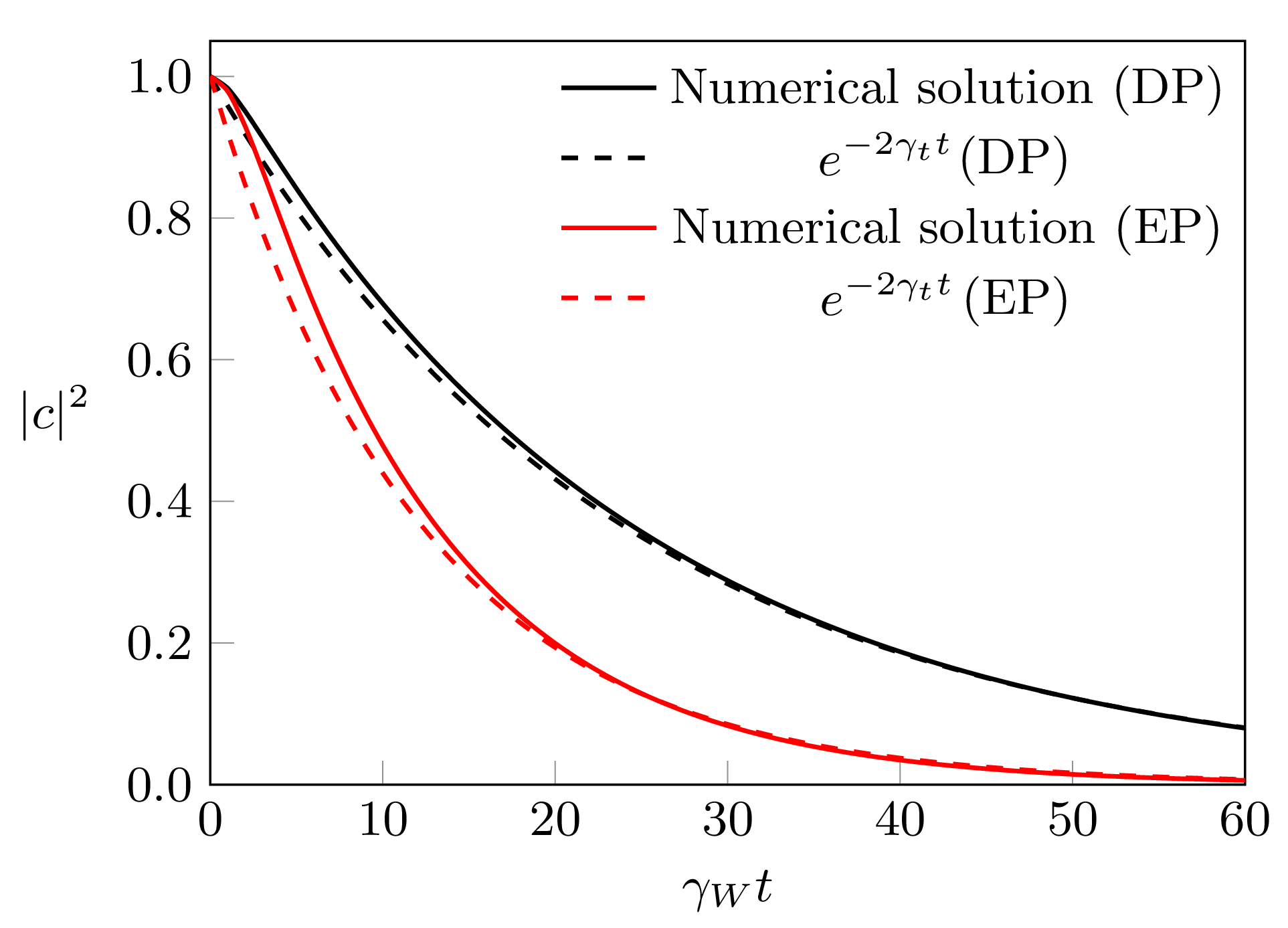}
		\caption{Comparison between the spontaneous emission (SE) decay rates as obtained by an exact numerical solution of Eq.\:(\ref{Eq:CMT}) (solid lines) and the analytic expression of Eq.\:(\ref{Eq:Decay_Appendix}) (dashed lines)  when the system operates at a diabolic point (DP) ($|r|=0$) and when the system operates at an exceptional point (EP) with maximum $\eta$ enhancement ($|r|=1$ and $\Delta\phi=\pi$). The simulation parameters were taken to be $\omega_o=\omega_e$ , $\frac{J}{\gamma_W}=10^{-1}$, and $\frac{\gamma_e}{\gamma_W}=10^{-3}$. For these parameters, using the approximation $\gamma_t \approx \text{Im}\{\delta\}$ does not introduce any visible changes to the plots, which justifies neglecting $\gamma_e$ in this regime.}
		\label{fig-Decay_rate}
	\end{center}
\end{figure}

In general, within the Purcell regime, where $\gamma_e\ll J\ll \gamma_W$ and $\gamma_e\ll J^2/\gamma_W$ \cite{Bozhevolnyi:16}, one can use the approximation $\alpha \approx \gamma_W$ in the expression for $\delta$. Moreover, in the SE enhanced decay regime, we can neglect $\gamma_e$ compared with $\text{Im}\{\delta\}$. This is basically the procedure we followed in solving Eq.\:(\ref{Eq:CMT}) in the main text. In this case, the above formula indeed gives the same result obtained in the main text using the normalized emitted power, i.e., $2\text{Im}\{\delta\}=P_{\text{EP}}$, with the equality holding also when $|r|=0$, i.e., in the DP case. However, as we pointed out, in the SE decay suppression regime when $\text{Im}\{\delta\}=0$, the excited QE will relax to its ground state by emission into free space with rate $\gamma_e$.

Finally, we confirm the above results by plotting the function $e^{-2\gamma_t t}$ and the full numerical solution of Eq.\:(\ref{Eq:CMT}) on the same graph for different cases, as shown in Fig.\:\ref{fig-Decay_rate}. Clearly, a good agreement is observed for the relevant parameters listed in the figure caption.

\section{Numerical evaluation of the optical parameters of the structure}\label{Appendix: numerical evaluation}

\begin{figure}[!ht]
	\begin{center}
		\includegraphics[width=3.4in]{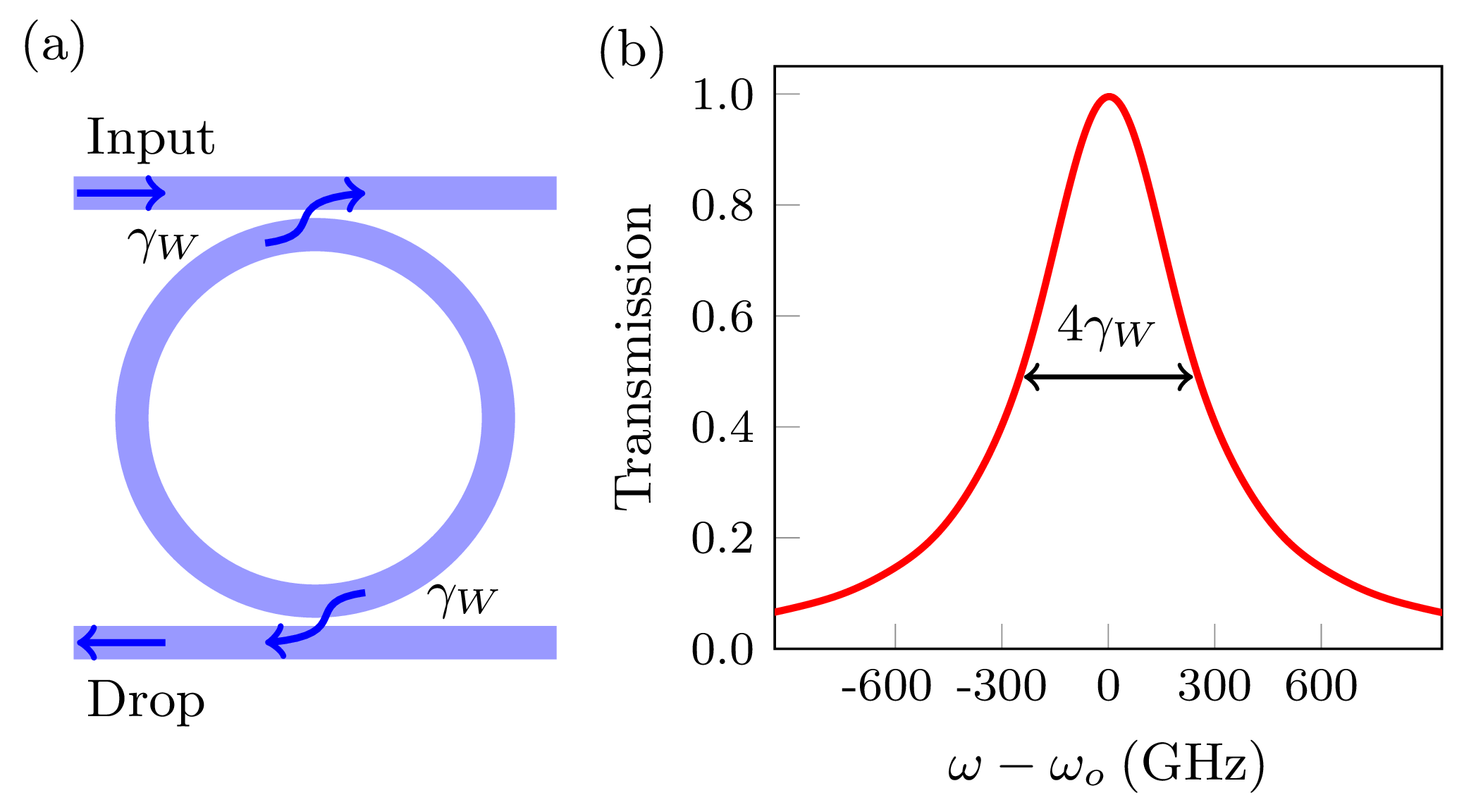}
		\caption{(a) An add-drop configuration is used to evaluate the decay rate from the resonator to the waveguide. (b) Optical power transmission as a function of frequency as obtained by using finite-difference time-domain (FDTD). Based on these simulations, we estimate that $4\gamma_W=496$ GHz.}
		\label{fig-AddDrop}
	\end{center}
\end{figure}

In the main text, we have compared the results obtained by using FDTD to those estimated using the coupled mode theory (CMT) as represented by the Hamiltonian $\hat{H}$. To use this later approach, however, one must evaluate the essential optical parameters characterizing the system, such as the coupling between the microring resonator and the waveguide (together with the related optical loss from the ring to the waveguide port), the mirror complex reflectivity, as well as its transmission (and hence its optical absorption). In our paper, these parameters were computed numerically using FDTD. Here, we outline the details of these calculations.

\begin{figure}[t]
	\begin{center}
		\includegraphics[width=3.4in]{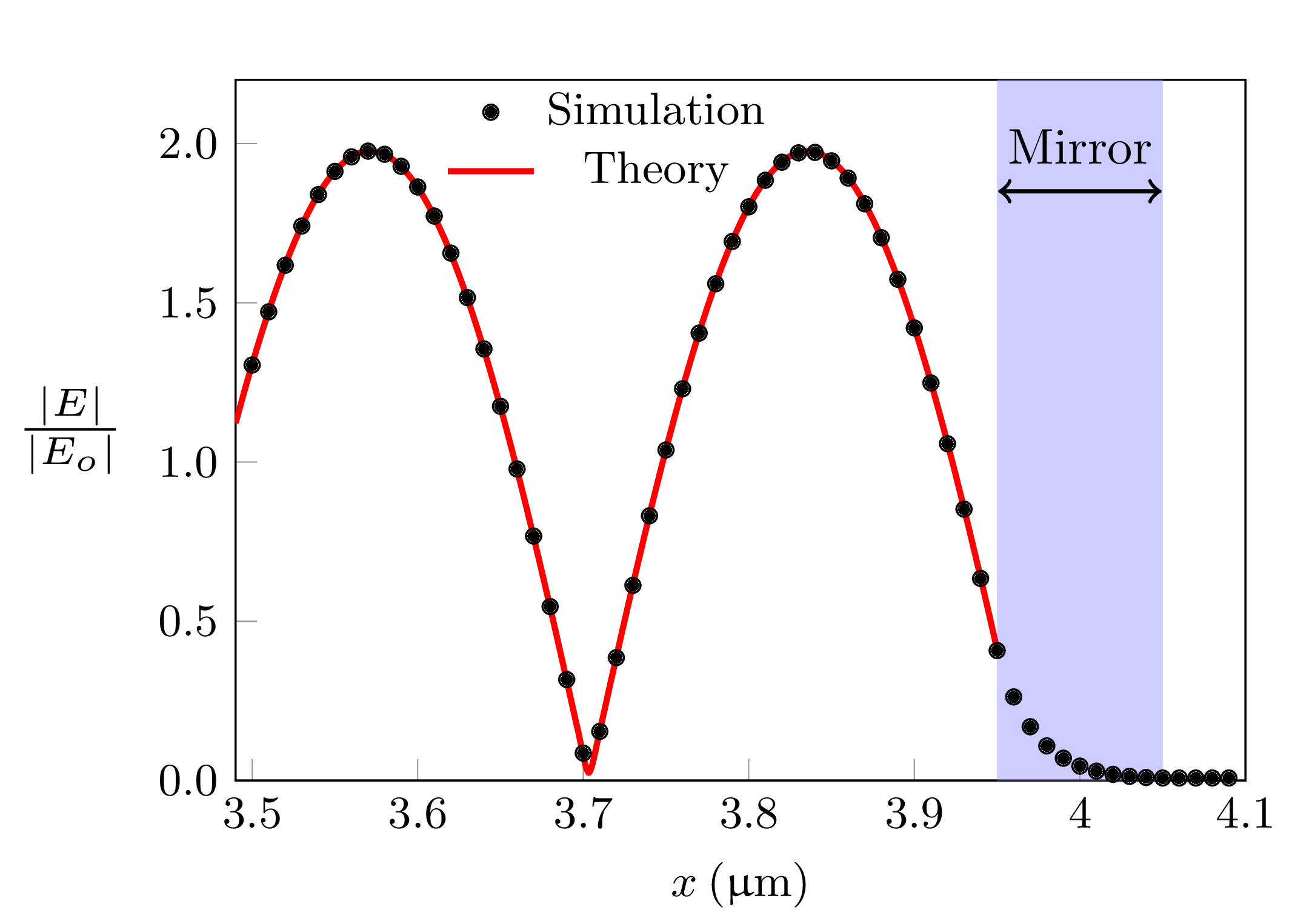}
		\caption{Absolute value of the electric field distribution when an optical mode is launched from the left side of the waveguide onto the mirror (in the absence of the resonator). Black dots are numerical data obtained from finite-difference time-domain (FDTD), and the red curve is the fit according to the expression of Eq.\:(\ref{Eq:standing_field}). The horizontal axis $x$ is the distance  measured from the position of waveguide-ring junction of the full geometry.}
		\label{fig-Mirror}
	\end{center}
\end{figure}

To extract the decay rate from the microring to the waveguide ($\gamma_W$), we consider an add-drop ring resonator filter geometry, as shown in Fig.\:\ref{fig-AddDrop}(a). The amplitude of the power transmission coefficient as a function of the input wave frequency is then obtained using FDTD [Fig.\:\ref{fig-AddDrop}(b)].   By noting that the full width at half maximum (FWHM) of the transmitted power is given by $4\gamma_W$, we estimate that, in our case, $\gamma_W = 124\:$GHz, which together with $\omega_o=1216 \: \text{THz}$ corresponds to a quality factor of $Q=4900$ for the all-pass structure.

Next, we calculate the optical parameters of the mirror by removing the ring resonator altogether and using FDTD simulations to evaluate the field reflection and transmission coefficient from the mirror. In these simulations, an incident optical mode having a free space wavelength of $\lambda_o=1549\:$nm is launched into the waveguide from the left. The absolute values of the reflection and transmission coefficient are easily obtained by numerically measuring the reflected and transmitted optical powers, respectively. Doing so gives $|r|=0.976$ and $|t|=0.008$. To evaluate the phase of the reflection coefficient, we plot the absolute value of the steady state electric field $|E|$ distributed at the incident wave side, as shown in Fig.\:\ref{fig-Mirror}, and use curve fitting based on the analytical expression
\begin{equation} \label{Eq:standing_field}
|E|=|E_o|\sqrt{1+|r|^2+2|r|\cos[2\beta (x-L)-\phi_r]},
\end{equation}
where here, $|E_o|$ is the incident wave amplitude, $|r|$ is the field amplitude reflection coefficient, $L=3.95 \ \mu \text{m}$ is the distance between the mirror and the waveguide-ring junction, $\phi_r$ is the reflection phase, and $\beta=\frac{2\pi n_\text{eff}}{\lambda}=11.891\: \mu \text{m}^{-1}$. Based on these values, we find that $\phi_r=3.56$ gives the best fit between Eq.\:(\ref{Eq:standing_field}) and the data in Fig.\:\ref{fig-Mirror}. However, to obtain the best match between CMT and FDTD for the full structure, we used $\phi_r=3.61$, which is an accepted discrepancy given the approximate nature of CMT.

\begin{acknowledgments}
R.E.  acknowledges support from the Army Research Office (ARO; Grant No. W911NF-17-1-0481), National Science Foundation (NSF; Grant No. ECCS 1807552), and the Max Planck Institute for the Physics of Complex Systems.  S.K.O. acknowledges support from ARO (Grant No. W911NF-18-1-0043), NSF (Grant No. ECCS 1807485), and Air Force Office of Scientific Research  (Award No. FA9550-18-1-0235).\\
\end{acknowledgments}

Q.Z. and A.H. contributed equally to this work.

\bibliography{Reference}

\end{document}